\documentclass[english,twocolumn,prl,amssymb,amsmath]{revtex4}
\usepackage[latin9]{inputenc}
\usepackage{graphicx,color}
\usepackage{hyperref}
\usepackage{babel}
\usepackage{booktabs}
\usepackage[normalem]{ulem}
\usepackage{dsfont}
\usepackage[percent]{overpic}

\usepackage{stackengine}
\stackMath

\newcommand{\be}{\begin{equation}}
\newcommand{\ee}{\end{equation}}

\definecolor{mygreen}{rgb}{0,0.5,0}
\definecolor{myblue}{rgb}{0,0,0.75}
\definecolor{mymagenta}{cmyk}{0,1,0,0.12}

\newcommand{\pd}{{\phantom{\dagger}}}

\begin{document}

\title{Unpaired Weyl nodes from Long-Ranged Interactions: Fate of Quantum Anomalies}


\author{Tobias Meng}
\affiliation{Institute of Theoretical Physics, Technische Universit\"at Dresden, 01062 Dresden, Germany}	
\author{Jan Carl Budich}
\affiliation{Institute of Theoretical Physics, Technische Universit\"at Dresden, 01062 Dresden, Germany}	
\date{\today}

\begin{abstract}
We study the effect of long-ranged interactions on Weyl semimetals. Such interactions can give rise to unpaired Weyl nodes, which we demonstrate by explicitly constructing a system with just a single node -- a situation that is fundamentally forbidden by fermion doubling in non-interacting band structures. Adding a magnetic field, we investigate the fate of the chiral anomaly. Remarkably, as long as a system exhibits a single Weyl node in the absence of magnetic fields, arbitrarily weak fields qualitatively restore the lowest Landau level structure of a non-interacting Weyl semimetal. This underlines the universality of the chiral anomaly in the context of Weyl semimetals. We furthermore demonstrate how the topologically protected Fermi-arc surface states are modified by long-ranged interactions.    
\end{abstract}

\date{\today}

\maketitle

\emph{Introduction. } Among the most intriguing phenomena associated with topological phases are their characteristic transport properties. 
A paradigmatic example is provided by the conducting surface states of topological insulators, in which the degrees of freedom expected for an ordinary metal with the same symmetries decompose into two halves that are spatially separated over opposite surfaces of the system \cite{HasanKane,XLReview}. For the more recently discovered Weyl semimetals (WSMs) \cite{volovik_book,WSM-exp-1,WSM-exp-2,WSM-exp-3,WSM-exp-4,WSM-exp-5,WSM-exp-6,WSM-exp-7,turner_review,armitage_review}, a spin-degenerate {\textit{bulk}} semimetal splits in reciprocal space into a pair of topologically stable Weyl nodes, giving rise to bulk transport anomalies such as the chiral anomaly or the mixed axial-gravitational anomaly \cite{adler,bell_jackiw,nielsen_ninomiya1,turner_review,landsteiner_anomalies_review,burkov_review,gooth_anomaly}. These anomalies are examples of quantum anomalies, the breaking of a classical conservation law due to quantum fluctuations \cite{bilal_lecture_anomalies}. Specifically, the chiral anomaly is the non-conservation of the number of electrons close to an individual Weyl node despite the presence of a $U(1)$-symmetry in its low-energy Weyl Hamiltonian. A crucial difference between WSMs and the aforementioned surface states of topological insulators is the fact that the topological stability of Weyl nodes is rooted in their separation in momentum space rather than real space \cite{armitage_review}. Hence, (lattice) momentum conservation is more important to the stability of WSMs than for topological insulators, the robustness of which against both disorder and interaction-induced scattering is well-known \cite{HasanKane}. 

\begin{figure}%
\quad\quad
\begin{overpic}[height=0.28\columnwidth]{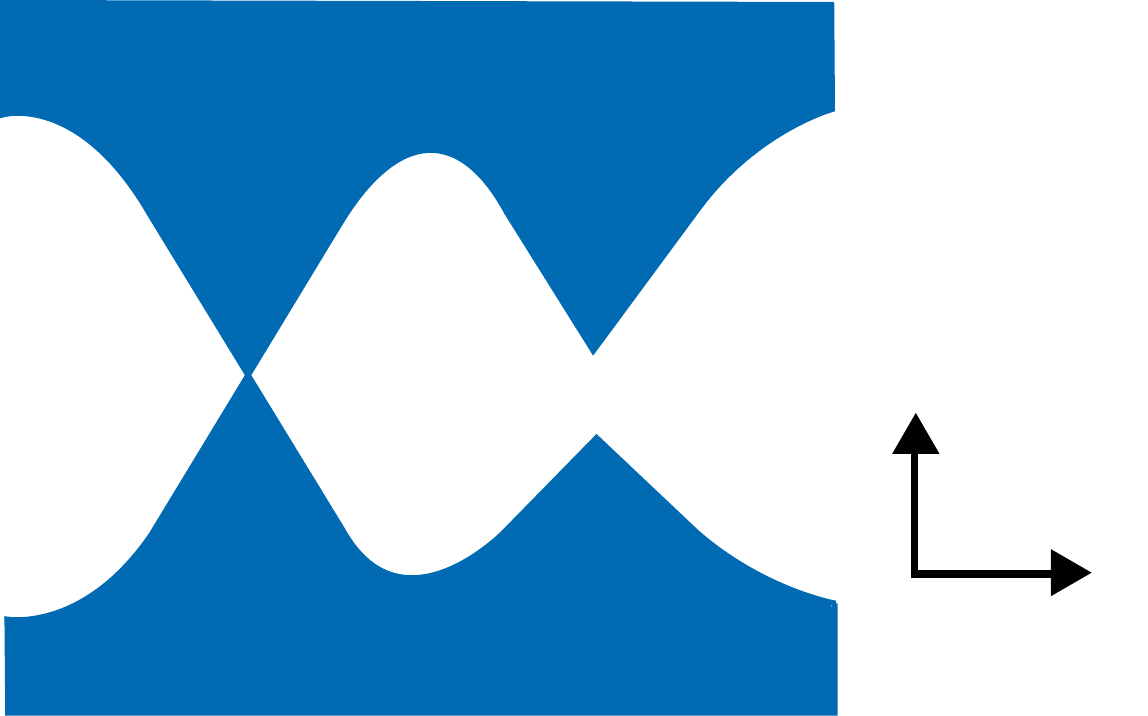}
 \put (77.5,29) {$\omega$}
  \put (97,10) {$k_z$}
 \put (-21,56) { (a)}
  \put (29,34) {$\scriptstyle B_z=0$}
    \put (26,26) {$\scriptstyle U(\mathbf{k})\neq0$}
 \end{overpic}
 \quad\quad
 \begin{overpic}[height=0.28\columnwidth]{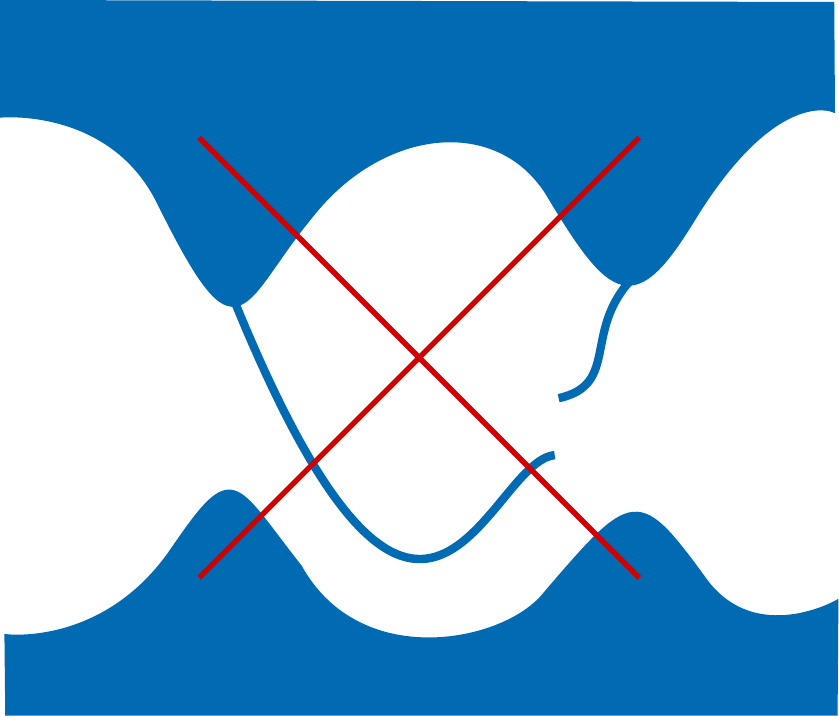}
 \put (-31.5,77) { (d)}
   \put (39.5,50) {$\scriptstyle B_z\neq0$}
    \put (34.5,39) {$\scriptstyle U(\mathbf{k})\neq0$}
 \end{overpic}
\\\hrulefill\\[0.2cm]
\begin{minipage}{0.4\columnwidth}\begin{center}
\begin{overpic}[width=\columnwidth]{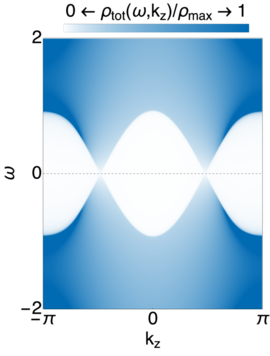}
 \put (-5,92)  { (b)}
 \put (35.5,53) {$\scriptstyle B_z=0$}
  \put (33,46) {$\scriptstyle U(\mathbf{k})=0$}
 \end{overpic}
\end{center}\end{minipage}
\quad\quad\quad\begin{minipage}{0.4\columnwidth}\begin{center}
\begin{overpic}[width=\columnwidth]{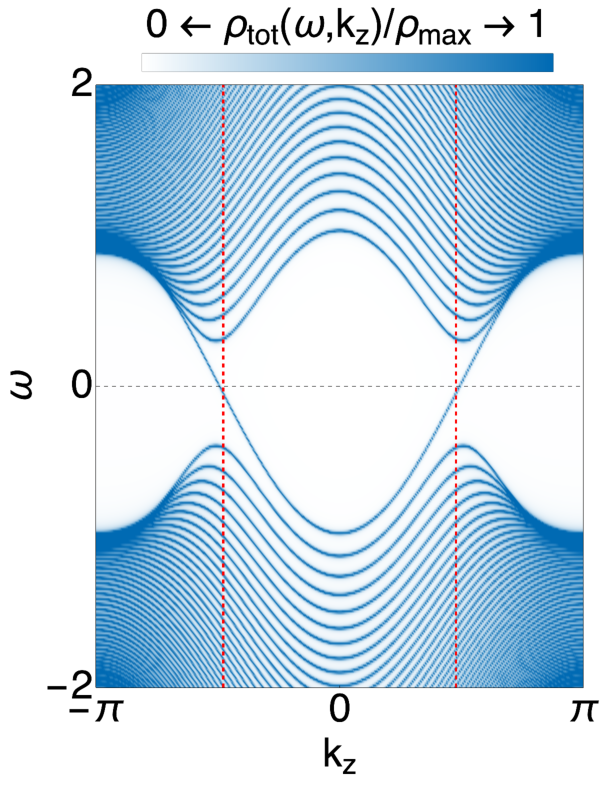}
 \put (-5,92) { (e)}
 \put (35.5,53) {$\scriptstyle B_z\neq0$}
  \put (33,46)  {$\scriptstyle U(\mathbf{k})=0$}
 \end{overpic}
\end{center}\end{minipage}\\[0.3cm]
\begin{minipage}{0.4\columnwidth}\begin{center}
\begin{overpic}[width=\columnwidth]{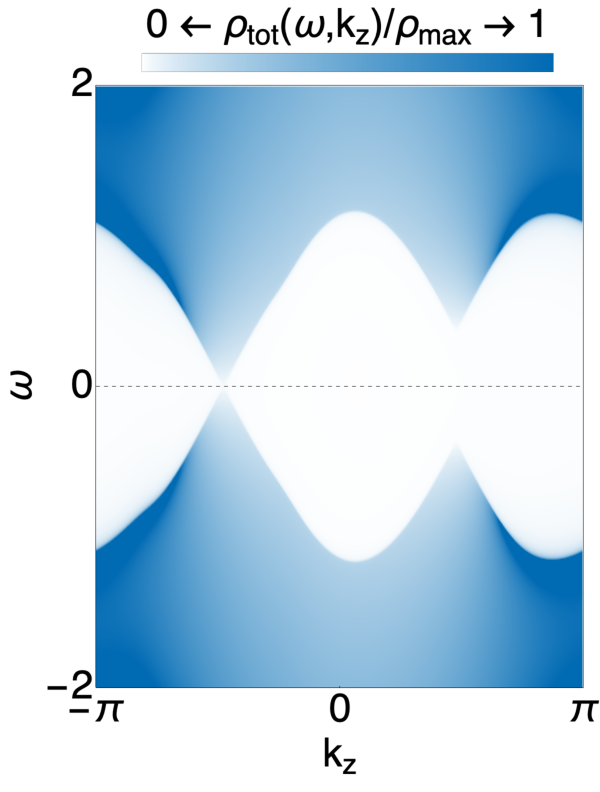}
 \put (-5,92)  { (c)}
 \put (35.5,53) {$\scriptstyle B_z=0$}
  \put (33,46)  {$\scriptstyle U(\mathbf{k}_\textrm{Weyl})\neq0$}
 \end{overpic}
\end{center}\end{minipage}
\quad\quad\quad\begin{minipage}{0.4\columnwidth}\begin{center}
\begin{overpic}[width=\columnwidth]{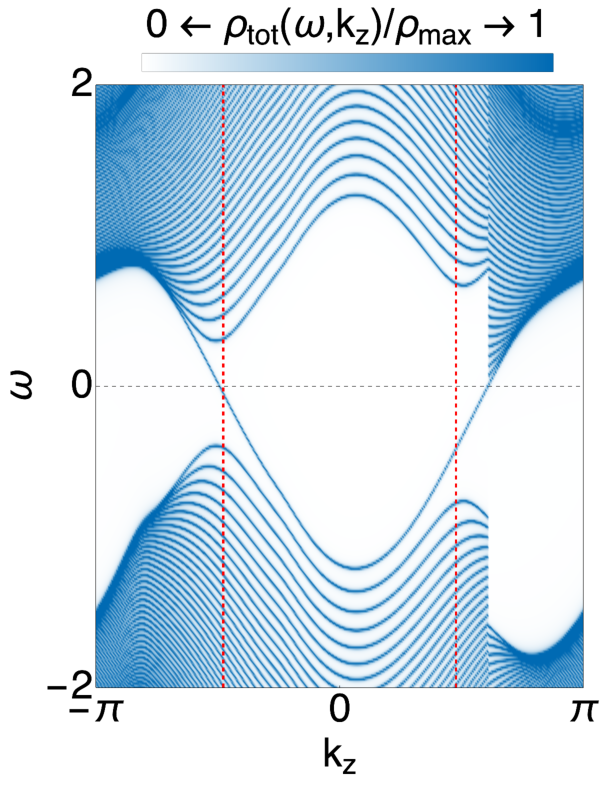}
 \put (-5,92)  { (f)}
 \put (35.5,53) {$\scriptstyle B_z\neq0$}
  \put (33,46)  {$\scriptstyle U(\mathbf{k}_\textrm{Weyl})\hspace{0.15cm}\neq0$}
 \end{overpic}
\end{center}\end{minipage}%
\caption{Total density of states of a Weyl semimetal with a momentum-local interaction. We plot $\rho_{\rm tot}(\omega,k_z)=\sum_{k_x,k_y}(-1/\pi)\,\sum_m\,\text{Im}\,G^R_m(\omega,\mathbf{k})$, where $G^R_m(\omega,\mathbf{k})$ is the retarded Green's function of the $m^{\rm th}$ eigenstate, see Eqs.~\eqref{eq:bulk_gf1} and \eqref{eq:gfm} \cite{footPlot}. Panel (a)-(c): total density of states rescaled by its maximum value $\rho_\textrm{max}$ as a function of frequency $\omega$ and lattice momentum $k_z$ in the absence of a magnetic field.  Panel (d)-(f): total density of states in the presence of a magnetic field $B_z$ in $z$-direction. Top row: Panel (a) schematically illustrates a single Weyl node in the presence of a momentum-local interaction. Such interactions, however, \textit{cannot} open a partial gap in the dispersion of zeroth Landau levels as sketched in (d). Central row: Numerical data for the model Hamiltonian (\ref{eq:ham}) in the absence of interactions ($U(\mathbf{k})=0$). Bottom row: Numerical data for the model Hamiltonian (\ref{eq:ham}) with interactions (in particular $U(\mathbf{k}_\textrm{Weyl})\ne0$), quantitatively corroborating the schematic scenarios in the top row. When the zeroth Landau level crosses zero in the interacting regime, the other energy levels show a discontinuous jump since the number of occupied states, and hence the interaction energy, changes abruptly. The magnetic unit cell in $x$-direction contains $100$ sites in panels panels (e) and (f).  The interaction parameters in panels (c) and (f) are $U_0=2,~\delta=2.5$. The magnetic field in panels (e) and (f) is $B_z=2\pi/100$. In all plots $\lambda=1.5$, $k_{\rm Weyl}=1.5$ \cite{more_plots}.
}%
\label{fig:fig1}%
\end{figure}
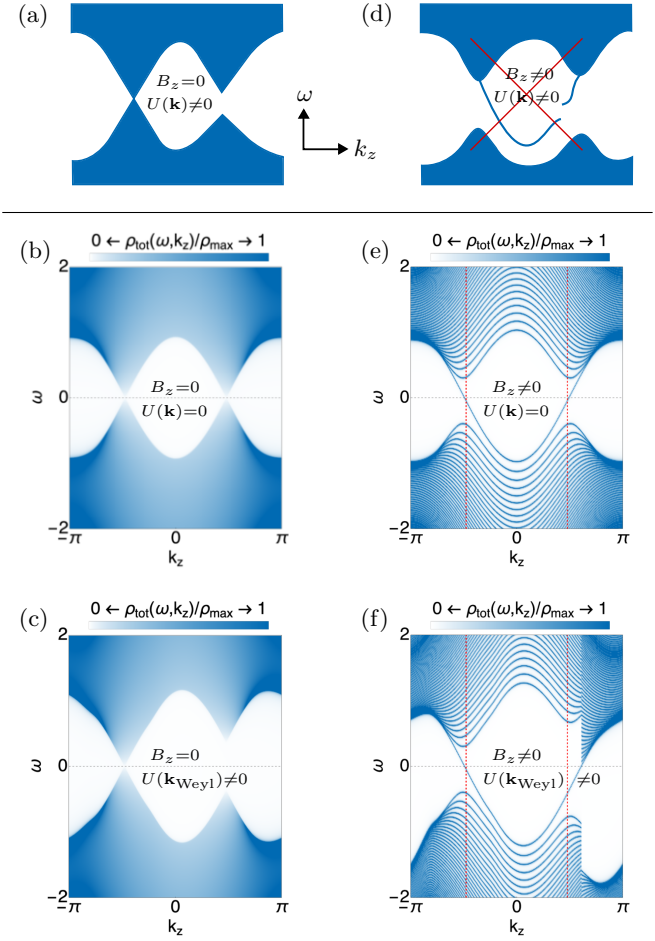

\begin{figure}%
\begin{minipage}{0.4\columnwidth}\begin{center}
\begin{overpic}[width=\columnwidth]{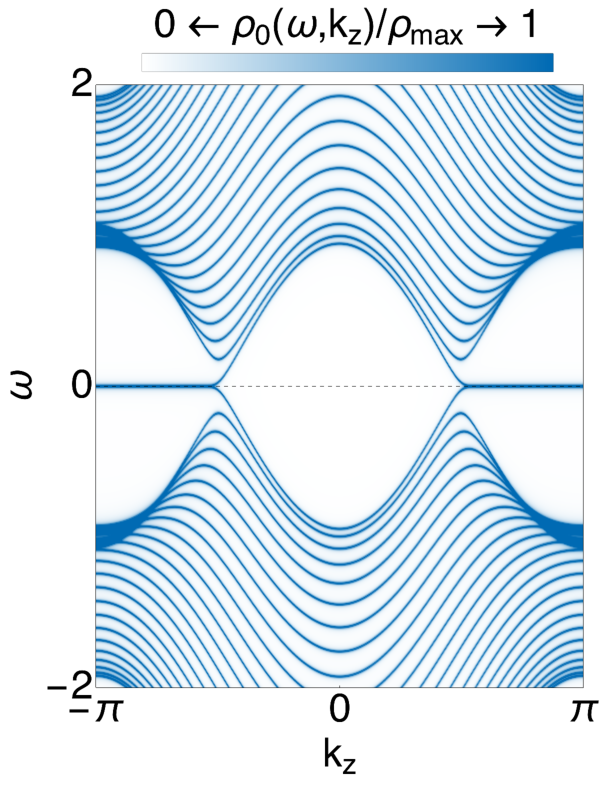}
 \put (-5,92)  { (a)}
 \put (35.5,53) {$\scriptstyle B_z=0$}
  \put (33,46) {$\scriptstyle U(\mathbf{k})=0$}
 \end{overpic}
\end{center}\end{minipage}\quad\quad\quad\begin{minipage}{0.4\columnwidth}\begin{center}
\begin{overpic}[width=\columnwidth]{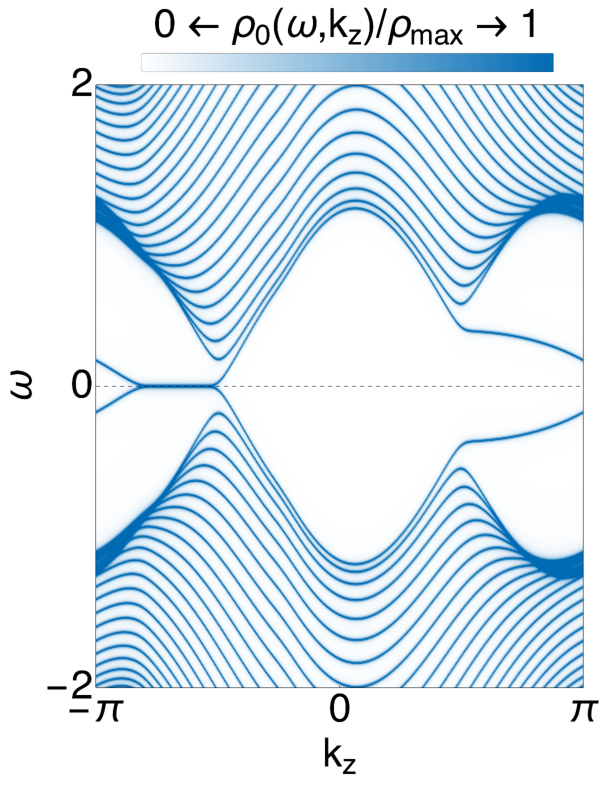}
 \put (-5,92)  { (b)}
 \put (35.5,53) {$\scriptstyle B_z=0$}
  \put (33,46) {$\scriptstyle U(\mathbf{k})\neq0$}
 \end{overpic}
\end{center}\end{minipage}
\caption{Total density of states $\rho_{k_y=0}(\omega,k_z)$ rescaled by its maximum value $\rho_\textrm{max}$ of a slab with open boundary conditions (i.e.~with surfaces) in zero magnetic field. The interaction is $U_0=0$ in panel (a) and $U_0=2$ in panel (b). The slab has 25 layers in $x$-direction, $\lambda=1.5$, $k_{\rm Weyl}=1.5$, and $\delta=2.5$ in all plots \cite{more_plots}.}%
\label{fig:nofield_surf_bands}%
\end{figure}
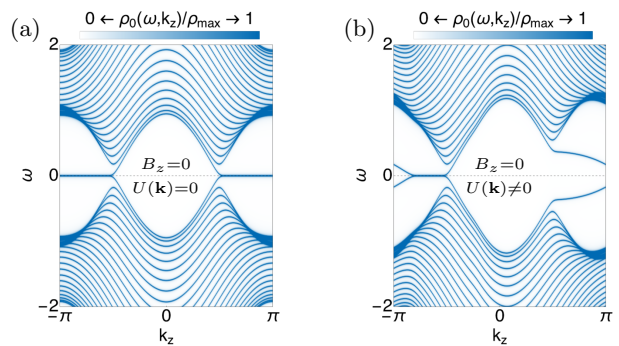

The purpose of this work is to investigate how WSMs are influenced by {\textit{long-ranged}} interactions that, due to their short-ranged nature in momentum space, do not couple different Weyl nodes in reciprocal space. In this scenario, we show that {\textit{effectively momentum dependent interactions}} can lead to the occurrence of single (unpaired) Weyl nodes, a situation that is fundamentally forbidden by the Nielsen-Ninomiya fermion doubling theorem in non-interacting systems \cite{nielsen_ninomiya1,nielsen_ninomiya2,nielsen_ninomiya3,nielsen_ninomiya4}, and that cannot be achieved when interactions have only a perturbative effect \cite{carlstrom_18}. The topological stability of WSMs can be reconciled with the existence of unpaired Weyl nodes due to the occurrence of characteristic Green's function zeros that replace the nodal points in the spectrum (related physics have been discussed in the context of interacting topological insulators \cite{volovik_zeros,GurarieInteractingTopology,ZhangPRB2012}). Focusing on an exactly solvable model with interactions that are local in momentum space \cite{MorimotoMottWeyl}, we study the fate of the chiral anomaly [see Fig.~\ref{fig:fig1}] and of the topologically protected Fermi-arc surface states [see Fig.~\ref{fig:nofield_surf_bands}] in strongly correlated systems with unpaired Weyl nodes. Quite remarkably, the characteristic lowest Landau level structure entailing the chiral anomaly with global electron number conservation is found to persist even in the presence of such quite extreme long-ranged interactions. Furthermore, we argue how interactions that single out an individual Weyl node can be realized microscopically in multi-orbital models.

\emph{Single Weyl nodes and topology at zero field.}
The defining feature of WSMs are linear band touching points described by the Weyl Hamiltonian \cite{weyl_29,herring_37,volovik_book}, whose first-quantized form is 
\begin{equation}
\hat{H}_{\rm Weyl}^\pm=\pm v_F\,\hat{\mathbf{\sigma}}\cdot\hat{\mathbf{q}},
\label{eqn:WeylHam}
\end{equation}
where $\hat{\mathbf{\sigma}}$ is the vector of standard Pauli matrices denoting a (pseudo-) spin-$1/2$ operator, $v_F$ stands for the Fermi velocity, and $\hat{\mathbf{q}}$ measures the three-dimensional momentum relative to the nodal point. The Nielsen-Ninomiya fermion doubling theorem  \cite{nielsen_ninomiya1,nielsen_ninomiya2,nielsen_ninomiya3,nielsen_ninomiya4} states that in a solid, these band touching points (Weyl nodes) always occur in pairs of opposite chirality as indicated by the index $\pm$ in Eq. (\ref{eqn:WeylHam}). This theorem, however, relies on a set of conditions which most importantly include the short-ranged nature of the underlying Hamiltonian. Here, we are instead considering the effect of long-ranged interactions on WSMs. We start by studying the exactly solvable Hamiltonian 

\begin{align}
&H=\sum_{\bf k}\left(\Psi_{\bf k}^\dagger \, \left [{\bf h}({\bf k})\cdot{\bf \hat \sigma}\right] \,\Psi_{\bf k}^\pd + \frac{U(\mathbf{k})}{2}\left(\Psi_{\bf k}^\dagger \Psi_{\bf k}^\pd-1\right)^2 \right), \label{eq:ham}
\end{align}
where ${\bf h}({\bf k})=\left(\sin(k_x),\sin(k_y),h_z(\bf k)\right)$ with $$h_z({\bf k})=\cos(k_z)-\cos(k_{\textrm{Weyl}})+\lambda(2-\cos(k_x)-\cos(k_y))$$ describes a minimal lattice model of a Weyl semimetal, and $\Psi_{\bf k}^\dagger=(c_{\uparrow{\bf k}}^\dagger,c_{\downarrow{\bf k}}^\dagger)$ denotes the spinor of creation operators of electrons of (pseudo\nolinebreak-) spin $\sigma=\uparrow, \downarrow$. As a crucial difference to earlier studies of long-ranged interactions \cite{MorimotoMottWeyl,SharmaPRB}, the interaction strength $U(\mathbf{k})$ is momentum-local but allowed to vary with momentum. The momentum-locality of the interaction makes it infinitely ranged in real space, and amounts to the constraint of not changing the center of mass of the interacting particles over a scattering process \cite{HatsugaiKohmoto}. For $U(\mathbf{k})=0$ and $\lambda >1$, this Hamiltonian features two Weyl nodes [see Eq.~\eqref{eqn:WeylHam}] at momentum $\pm\mathbf{k}_{\rm Weyl}^T=(0,0,\pm k_{\rm Weyl})$. These nodes for example show up in spectroscopic measurements of the bulk density of states [see Fig.~\ref{fig:fig1} for an illustration].

We now switch on the interaction $U(\mathbf{k})$ only in the vicinity of one of the nodes. To this end, we introduce a smooth function describing a bump of width $\delta$ around $k_0$, defined as $b(k,k_0,\delta)=\text{exp}\{-\delta^2/(\delta^2-(k-k_0)^2)\}$ for $k\in[k_0-\delta,k_0+\delta]$, and zero elsewhere \cite{FootBump,supplement}. The interaction is then chosen as $U(\mathbf{k}) = U_0\,b(k_z,k_{\rm Weyl},\delta)$ with an appropriate momentum width $\delta$. Since the Hamiltonian
(\ref{eq:ham}) is local in momentum space, we can readily find the many-body eigenstates and -energies by exact diagonalization of the Hamiltonian for every fixed momentum $\mathbf{k}$. In the absence of a magnetic field, the Hilbert space associated with a given momentum is only four-dimensional, and the Hamiltonian is similar to the one of a spinful quantum dot, for which it is known that an interaction leads to Coulomb blockade, the quantum dot analogue of a Mott gap \cite{bruus_flensberg}.

To demonstrate how such a momentum-local interaction can give rise to single Weyl nodes, we calculate the Green's function $G$ in Lehman representation after diagonalizing the Hamiltonian for fixed $\mathbf{k}$. Away from the Weyl nodes, it reads as
 
 \begin{align}
 G^R(\omega,\mathbf{k})&= \left(\omega-\mathbf{h}_{\rm eff}(\mathbf{k})\cdot\mathbf{\sigma}+i\eta\right)^{-1}\label{eq:bulk_gf1}
 \end{align}
with $\mathbf{h}_{\rm eff}(\mathbf{k})={\bf h}({\bf k})\left(1+\frac{U(\mathbf{k})}{2|{\bf h}({\bf k})|}\right)$, while it is given by
\begin{align}
&G^R(\omega,\pm \mathbf{k}_{\textrm{Weyl}})= \nonumber\\&\mathds{1}\left(\frac{1/2}{\omega+\frac{U(\pm \mathbf{k}_{\textrm{Weyl}})}{2}+i\eta}+ \frac{1/2}{\omega-\frac{U(\pm \mathbf{k}_{\textrm{Weyl}})}{2}+i\eta}\right)
\label{eqn:WeylGreen}
\end{align}
at the Weyl nodes (with $\eta\to0^+$). The system's excitation energies (the poles of its Green's function) are simply shifted by an energy $\Delta E=\pm U(\mathbf{k})/2$ due to interactions. This is reflected in the density of states, $\rho(\omega,\mathbf{k})=(-1/\pi)\,\sum_m\,\text{Im}\,G^R_m(\omega,\mathbf{k})$, by a clear Mott-gap visible around the node at $\mathbf{k}=+\mathbf{k}_{\rm Weyl}$ as shown in Fig.~\ref{fig:fig1} \cite{MorimotoMottWeyl} (where $G^R_m(\omega,\mathbf{k})$ is the retarded Green's function of the $m^{\rm th}$ eigenstate, see Eqs.~\eqref{eq:bulk_gf1}, and \eqref{eq:gfm} below). We have thus constructed a minimal exactly solvable model for an interacting system exhibiting a \textit{single unpaired Weyl node}. Below, we will discuss how the crucial momentum-dependence of the interactions that is necessary for singling out an individual Weyl node can be realized in a microscopic model with four bands, i.e. with an additional orbital degree of freedom.

We now address how this anomalous spectrum can be reconciled with the topological stability of Weyl nodes. In a modern language of topology, Weyl nodes are associated with quantized monopoles of the Berry curvature. Hence, the Berry flux through $(k_x,k_y)$-planes at fixed $k_z$, which equals $2\pi$ times the Chern number of those planes, changes at the Weyl nodes, i.e.~at $k_z=\pm k_{\textrm{Weyl}}$. Due to the periodicity of $k_z$ in the first Brillouin zone, every monopole then must be compensated by an oppositely charged antagonist at which the change in the Chern number is reversed. This provides an intuitive picture of the aforementioned fermion doubling theorem. For interacting systems, Chern numbers $C$ can be defined in terms of single-particle Green's functions $G$ \cite{RedlichGF, SoGF, IshikawaGF, VolovikGF, GurarieInteractingTopology}. Explicitly, $C$ reads

\begin{align}
C&=\frac{\epsilon_{\alpha\beta\gamma}}{6}\int_{-\infty}^\infty\,d\omega \int \frac{d^2k}{(2\pi)^2}\nonumber\\
&\times {\rm tr}\left\{G^{-1}\partial_{k_{\alpha}} G G^{-1}\partial_{k_{\beta}} G G^{-1}\partial_{k_{\gamma}} G\right\},
\end{align}
where $\alpha,\beta,\gamma\in\{0,1,2\}$ (summation is understood), $k_0=\omega$, $\epsilon_{\alpha\beta\gamma}$ is the totally antisymmetric tensor, and tr indicates the trace over the Green's function matrix structure, while the momentum integral extends over a plane in momentum space. Given the expression of the Green's function in Eq.~\eqref{eq:bulk_gf1}, and the fact that the Berry curvature only depends on $\mathbf{n}(\mathbf{k})=\mathbf{h}_{\rm eff}(\mathbf{k}) / |\mathbf{h}_{\rm eff}(\mathbf{k})| = \mathbf{h}(\mathbf{k})/|\mathbf{h}(\mathbf{k})|$, one realizes that the Chern numbers in the interacting system at momenta away from the Weyl nodes are identical to ones of the non-interacting system \cite{MorimotoMottWeyl}. This in particular implies that the Chern number changes between $C=0$ and $C=1$ at the momenta of \textit{both} the gapless \textit{and} the gapped Weyl node. Such an ``unwinding'' of a topological defect without a gap closing is fundamentally forbidden for non-interacting systems, but allowed in the presence of interactions. Namely, while Chern numbers can only change via poles of the Green's function at zero energy in non-interacting systems, the presence of correlations allows for zeros of the Green's function at zero frequency which can also change the value of topological invariants  \cite{volovik_zeros,GurarieInteractingTopology}. Eq.~\eqref{eqn:WeylGreen} shows that this is indeed what happens at the gapped Weyl node, $G^R(0,\mathbf k_{\rm Weyl})=0$. On a related note, we stress that strong interactions can in principle also annihilate the aforementioned topologically protected metallic surface states via the formation of new types of gapped surface topological phases \cite{SenthilPRB2013,QiJSM2013,SenthilScience2014}. Returning to the Nielsen-Ninomiya theorem, the requirement for Weyl nodes to appear in pairs is relinquished in general interacting systems by the possibility of attaching monopole charges of the Berry curvature not only to Weyl nodes but also to zeros of the Green's function. 

\emph{Chiral anomaly at finite magnetic field field.}
One of the main hallmarks of a Weyl semimetal is the chiral anomaly of its Weyl nodes, i.e. is the non-conservation of the number of low-energy electrons close to a given Weyl node in the presence of electromagnetic fields \cite{nielsen_ninomiya1,turner_review,armitage_review,landsteiner_anomalies_review,burkov_review}. Originally discussed in high-energy field theories \cite{adler,bell_jackiw}, the chiral anomaly can be cast into a simple solid state language: in the presence of a magnetic field, a Weyl node decomposes into Landau levels that disperse parallel to the field. The zeroth Landau level is special in that it is gapless and linearly dispersing. When an electric field is applied parallel to the magnetic field, the momentum of the low-energy electrons parallel to the field increases due to the Coulomb force $\mathbf{F}=e\,\mathbf{E}=\dot{\mathbf{k}}$, which pumps electrons between the Weyl nodes. While the total number of low-energy electrons is constant as expected, the number of low-energy electrons close to a given node is not conserved.%

This naturally raises the question of how the chiral anomaly manifests in our present interacting Weyl semimetal with a single unpaired Weyl node. To understand the Landau level structure, and from there the fate of the chiral anomaly, two observations are important. First, the Landau level structure of the remaining gapless Weyl node at $\mathbf{k}=-\mathbf{k}_{\rm Weyl}$ is unchanged, and this node hence still exhibits the anomalous non-conservation of its low-energy electron number in electromagnetic fields. The physically required conservation of the total electron number then implies that there must be a second linearly dispersing Landau level crossing zero energy with opposite velocity somewhere else in the Brillouin zone, unless particle number conservation is restored at the surfaces \cite{pseudo_fields,Weyl_sc_anomaly}. At the same time, it is clear that an infinitesimally small magnetic field cannot close the finite gap at the gapped Weyl point $\mathbf{k}=\mathbf{k}_{\rm Weyl}$. As a consequence, the additional chiral level is can only cross zero energy at a momentum away from the gapped Weyl node. 

To test this intuition with exact calculations, we modify the lattice Hamiltonian in Eq.~\eqref{eq:ham} so as to include a magnetic field in $z$-direction by means of the Peierls substitution, and perform numerical simulations on a system with a finite magnetic unit cell (rational magnetic flux per unit cell). We work in a Landau gauge in which $k_y$ is still a good quantum number. In $x$-direction, the enlarged periodicity due to the magnetic unit cell implies that the original momentum $k_x$ is replaced by a combination of a new magnetic lattice momentum $k_x$ defined in the reduced Brillouin zone and a Landau level index $n$. The interaction Hamiltonian is still taken to couple all electrons with the same $\mathbf{k}$, $\sum_{\mathbf{k}}\frac{U(\mathbf{k})}{2}\left( \sum_{\sigma}\left(c_{\sigma \bf k}^\dagger c_{\sigma\bf k}^\pd-1/2\right)\right)^2  \to \sum_{\mathbf{k}}\frac{U(k_z)}{2}\left(\sum_{n}\left(c_{\mathbf{k}n}^\dagger c_{\mathbf{k}n}^\pd-1/2\right)\right)^2$. In Fig.~\ref{fig:fig1}, we show how the correlated WSM with an unpaired Weyl node is affected by a weak magnetic field. The chiral zeroth Landau level structure, associated with the two Weyl nodes in the absence of interactions, is continuously shifted in energy by interactions, thus pushing the zero-energy crossing from the second (gapped) Weyl node to larger $k_z$. More precisely, the energy dispersion of the Landau level crossing zero twice is unchanged compared to the non-interacting case up to a simple shift by $-U(\mathbf{k})/2$. This Landau level hence crosses zero energy at the momentum where the interaction energy becomes smaller than the magnetic energy. We furthermore find that this particular Landau level still exhibits a smooth dispersion in the presence of interactions, while all other Landau levels become discontinuous because of the number of occupied states, and hence the interaction energy, changes when the chiral Landau level is emptied upon crossing zero energy (the absence of jumps at zero magnetic field is due to the fact that no level crosses zero energy at momenta where the interaction is finite). This can be understood analytically by calculating the Green's function at zero temperature using Lehmann representation. For a given level $m$ that can be occupied or empty, the retarded Green's function reads

\begin{align}
G_{m,\mathbin{\stackanchor[8pt]{\scriptstyle\rm occupied}{\scriptstyle\rm empty}}}^R(\omega,\mathbf{k})=\frac{1}{\omega-\epsilon_{\mathbf{k},m}-U(\mathbf{k})\left(\Delta N_{\mathbf{k}}\mathbin{\stackanchor[6pt]{-}{+}}\frac{1}{2}\right)+i\eta},\label{eq:gfm}
\end{align}
where $\Delta N_{\mathbf{k}}=N_{{\rm GS}}(\mathbf{k})-N$ is the difference in the number of total occupied states at momentum $\mathbf{k}$ in the ground state, $N_{{\rm GS}}(\mathbf{k})$, and the number $N$ of layers in the unit cell (or slab). When a chiral level crosses zero energy, $\Delta N_{\mathbf{k}}$ changes by one. This results in a jump for all Green's functions except the one of the chiral level, in which the change in $\Delta N_{\mathbf{k}}$ is compensated by the change in the expression of the Green's function between occupied and empty levels \cite{supplement}.

\emph{Surface states and Fermi-arcs at zero field.} 
Finite systems can feature topologically protected surface states as a consequence of the so-called bulk-boundary correspondence \cite{GurarieInteractingTopology,qi_06,hatsugai_93,bulk_boundary_ryu,mong_bulk_bundary,essin_gurarie_bulk_boundary_gf}. For non-interacting systems, an intuitive explanation thereof is again provided by the fact that the change of topology from the system to vacuum requires a gap-closing, and hence metallic surface states. Planes at fixed $k_z$ with non-trivial Chern number, i.e. with $\lvert k_z\rvert >k_{\textrm{Weyl}}$, are thus associated with a zero-energy edge state. The surface Fermi line of these states is known to be an open line segment (the famous ``Fermi-arc'' \cite{wan_arc}) that ends in the projection of the bulk Weyl nodes onto the surface. 

To study how these surface states are affected by our momentum-local interaction, we now turn to a slab with open boundary conditions (the slab is taken to be finite in $x$-direction). Because the Chern numbers of planes with fixed $k_z$ away from the Weyl nodes are unchanged by the interactions, one might naively expect the topological zero-energy surface states to persist in the presence of interactions. In contrast to the predictions of Ref.~\cite{MorimotoMottWeyl}, we find that this is not the case. Instead, topology unwinds at the surface in precisely the same way as it does at the gapped Weyl node: the zero-energy poles of the surface Green's functions split into two poles of halved spectral weight at energy $\omega=\pm U(\mathbf{k})/2$, and a zero at zero energy, $G^R(0,\mathbf{}k_{\rm Weyl})=0$, which is reflected by a gap in the density of states. The other levels (the bulk states) are also pushed away from the Fermi level by $\pm U(\mathbf{k})/2$, as in the case of periodic boundary conditions. These findings are illustrated by the densities of states plotted in Fig.~\ref{fig:nofield_surf_bands}. Physically, the gap of the surface states at momenta $k_z$ with finite interactions is a consequence of the fact that a non-local interaction couples counter-propagating modes on opposite surfaces.

\emph{Other interaction profiles.} Thus far, we have focussed on an interaction profile with a single peak in momentum space. To check the generality of our results, we have performed additional simulations with other interaction profiles \cite{supplement}. Variations of the interaction range $\delta$ have no qualitative effect as long as only one node is gapless: $\delta$ simply marks the momentum range over which the levels in the density of states are shifted by $\pm U/2$. By contrast, if the interaction gaps both Weyl nodes, the zero energy crossings of the Landau levels generically disappear, and with them any signatures of the chiral anomaly. A residual signature of the gapped Weyl nodes may then be still found in open systems with an effectively momentum-dependent interaction that vanishes in some momentum range between the Weyl nodes: such a system shows Fermi-arc segments in the momentum range with vanishing interaction in the complete absence of bulk Weyl nodes \cite{supplement}. 

\emph{Concluding discussion.} Our calculations are based on two rather extreme assumptions: The locality of interactions in momentum space, and the explicit momentum-dependence of the interactions. While locality in momentum space is necessary to make the present 3D strongly correlated problem amenable to exact study, we expect all of the discussed low-energy physics to be qualitatively robust for interactions with a finite but limited range in momentum space so as to not couple different Weyl nodes. Short-ranged interactions that lead to transfer of momentum comparable to the separation of the Weyl nodes, however, will generically couple opposite Weyl nodes and thus lead to a breakdown of the limiting scenario captured by our exactly solvable model. Our assumption of explicitly momentum dependent interactions may not seem very natural but is crucial in order to single out individual Weyl nodes. To address this key remaining issue, we will now provide a concrete mechanism for achieving an {\emph{effectively momentum dependent interaction}} in a microscopic model. The basic idea is to use an orbital degree of freedom given by the orbitals $A$ and $B$ in addition to the spin $\sigma$. In such a four-band system, already at the level of a next-nearest-neighbor tight-binding model, a WSM with one Weyl node being accommodated in orbital $A$ only and the other one in orbital $B$ only can be readily achieved \cite{supplement}. In this situation, a different interaction between particles in orbitals $A$ and $B$ is clearly sufficient to obtain the desired effective momentum dependence. The limiting case of an interaction affecting only a single Weyl node then simply amounts to the situation where only particles in, say, orbital $B$ interact with each other. We note that in the context of synthetic material systems based on ultracold atoms in optical lattices \cite{Bloch2008}, where very recently first experimental signatures of 3D topological semimetals have been reported \cite{NodalLineSMColdAtoms}, such a state-dependent interaction can be achieved when using two different internal states of the atom with different low energy scattering properties as orbitals $A$ and $B$ \cite{yi_08,gerbier_10}. However, even in conventional materials, a certain orbital dependence of the interaction strength may be sufficient to give different relevance to correlations at opposite Weyl nodes.

Regarding the persistence of the chiral anomaly, quasi-one-dimensional chiral Landau levels can only be gapped when counter-propagating modes are coupled, while any coupling between modes moving into the same direction can merely shift their energies (and hence the momentum at which they cross zero energy, as we observe here). On a more technical note, this statement follows from a perturbative analysis in a bosonized version  of the effectively one-dimensional chiral low-energy modes representing the zeroth Landau levels. An interaction with limited range in momentum space (coupling only modes that move into the same direction) can either renormalize the quadratic part of the Luttinger liquid Hamiltonian, or at most give rise to a sine-Gordon term whose argument does not commute with itself at different positions. Neither case is associated with a gap opening. We thus expect the presently discussed physics to be realizable in multi-orbital systems with more realistic interactions. We note that a key assumption in our study is that the magnetic field is aligned along the axis separating the Weyl nodes (i.e. the $z$-axis in our model system). Deviating from this alignment already in non-interacting systems leads to a breakdown of the chiral anomaly \cite{Kim2017,Zhang2017}, and has similar implications in our interacting scenario.

Our results show that even in the exotic scenario of unpaired Weyl nodes due to long-ranged interactions, global electron number conservation strongly constrains the low-energy spectrum: as long as at least one gapless Weyl node remains, even an infinitesimal magnetic field qualitatively restores the low-energy Landau level structure of an ordinary WSM with a pair of nodes. This dramatic stability of the chiral zeroth Landau levels in systems with single Weyl nodes underlines the fundamental importance of quantum anomalies for the classification of topological states of matter \cite{ryu_anomalies}, even in the extreme limit of long-ranged interactions.

\begin{acknowledgments}We acknowledge helpful discussions with  E. J. Bergholtz, B. Sbierski, and M. Vojta. TM is supported by the Deutsche Forschungsgemeinschaft via the Emmy Noether Programme ME 4844/1-1 and through SFB 1143.
\end{acknowledgments}

\bibliographystyle{apsrev}

\begin{thebibliography}{10}
\bibitem{FootBump}
We backfold the bump function $b(k_z,k_0,\delta)$ into the first Brillouin zone if required to obtain a lattice periodic function.
\end{thebibliography}

\end{document}